\documentclass[twocolumn,aps,preprintnumbers,prl,showpacs,%
amsfonts,amsmath,amssymb,superscriptaddress,floatfix]{revtex4}

\usepackage[dvips]{graphicx}

\newcommand{\affA}{%
	Department of Applied Physics, School of Engineering, 
        The University of Tokyo,\\
	7-3-1 Hongo, Bunkyo-ku, Tokyo 113-8656, Japan}

\newcommand{\affB}{%
	CREST, Japan Science and Technology Agency, 
	1-9-9 Yaesu, Chuoh-ku, Tokyo 103-0028, Japan}

\newcommand{\affC}{Computer Science, University of York, York YO10 5DD, UK}

\begin{document}

\title{Demonstration of quantum telecloning of optical coherent states}

\date{\today}

\author{Satoshi Koike}

\affiliation{\affA}

\author{Hiroki Takahashi}

\author{Hidehiro Yonezawa}

\author{Nobuyuki Takei}

\affiliation{\affA}

\affiliation{\affB}

\author{Samuel L.\ Braunstein}

\affiliation{\affC}

\author{Takao Aoki}

\author{Akira Furusawa}

\affiliation{\affA}

\affiliation{\affB}

% \maketitle

\begin{abstract}
We demonstrate unconditional telecloning for the first time. In particular,
we symmetrically and unconditionally teleclone coherent states of light 
from one sender to two receivers, achieving a fidelity for each clone 
of $F = 0.58 \pm 0.01$, which surpasses the classical limit. This is a 
manipulation of a new type of multipartite entanglement whose nature is 
neither purely bipartite nor purely tripartite.
\end{abstract}


\maketitle

Quantum telecloning \cite{Murao99} is a quantum information protocol 
combining cloning and teleporting into a single new primitive. 
The protocol offers significant technical advantages over the two-step 
local cloning plus teleportation strategy. In particular, in the case of 
coherent state telecloning, only finite entanglement is required for 
generating remote clones with optimal fidelity \cite{vanLoock01}.

In fact, quantum telecloning generalizes quantum teleportation with 
multiple receivers \cite{Murao99}. In quantum teleportation, 
bipartite entanglement shared by two parties (Alice and Bob) enables 
them to teleport an unknown quantum state from Alice to Bob by 
communicating only through classical channels \cite{Bennett93}. If 
three parties (Alice, Bob, and Claire) share an appropriate tripartite 
entangled state, Alice is able to teleport an unknown quantum state 
to Bob and Claire simultaneously. This is called ``$1 \to 2$ quantum 
telecloning.'' More generally, quantum telecloning to an arbitrary number 
of receivers ($1 \to n$ quantum telecloning) can be performed by using 
multipartite entanglement.

The heart of quantum telecloning is the multipartite entanglement shared
among the sender and receivers. Without multipartite entanglement, only 
the corresponding two-step protocol is possible: first the sender makes 
clones locally \cite{Andersen05}, and then sends them to each receiver 
with bipartite quantum teleportation \cite{Furusawa98} (or vice versa, 
teleporting followed by local cloning). The two-step protocol would 
require maximal bipartite entanglement for optimal fidelity teleportation 
(which for continuous-variable teleportation corresponds to states with 
infinite energy). Surprisingly, continuous-variable telecloning of 
coherent states requires only finite squeezing to achieve the same 
optimal fidelity \cite{vanLoock01}. In fact, the level of squeezing 
needed for optimal telecloning of coherent states is close to the reach 
of current technology \cite{Polzik92}.

We demonstrate the quantum telecloning of coherent states.  
The ability to reliably manipulate coherent states is of particular 
relevance to cryptographic scenarios. Indeed, in all non-heralded 
or non-entangled quantum cryptosystems the states used are invariably 
coherent states.  More generally, experimental quantum telecloning provides 
us with another way of manipulating multipartite entanglement, a resource 
which plays an essential role in quantum computation and multiparty quantum 
communication. A related scheme for so-called partial teleportation involves
one local and one remote clone. This scheme was demonstrated for photonic
qubits \cite{zhao}, but could presumably be extended to coherent states.

% Coherent states are so far invariably used in non-heralded or non-entangled
% based implementations [REFS] --- simply because it has currently been so
% much easier to construct weak coherent states than single-photon states.

Here the quantum state to be telecloned is that of an electromagnetic
field mode. We use the Heisenberg picture to describe the evolution of the
quantum state. An electromagnetic field mode is represented by an
annihilation operator $\hat{a}$ whose real and imaginary parts ($\hat{a}
= \hat{x} + i \hat{p}$) corresponds to the position and momentum
quadrature-phase amplitude operators. These operators $\hat{x}$ and
$\hat{p}$ satisfy the commutation relation 
$[ \hat{x}, \hat{p} ] = \frac{i}{2}$ (in so-called photon-number units 
with $\hbar = \frac{1}{2}$) and can be treated 
as canonically conjugate variables. This continuous-variable approach 
has attracted much interest because of the relative ease of realization of
unconditional or deterministic quantum information processing \cite{RMP}.
Unconditional quantum teleportation was demonstrated for the first time
with this approach \cite{Furusawa98}, and various successful experiments
have been reported
\cite{Andersen05,Furusawa98,Bowen03,Zhang03,Li02,Jing03,Yonezawa04,Mizuno05}.

% Let us concentrate on $1 \to 2$ quantum telecloning of a coherent state
% input \cite{vanLoock01}. Quantum telecloning requires entanglement which 
% comprises both a bipartite and tripartite structure similar to the $W$ 
% state \cite{Duer00}, although it is not maximally entangled. We generate 
% this type of entanglement using two squeezed-vacuum states with a modest 
% amount of squeezing and a pair of 50/50 beam splitters. The scheme for 
% creating the tripartite entanglement for quantum telecloning 
% \cite{vanLoock01} is shown in the center of Fig.~\ref{fig:setup}. Two optical 

% parametric oscillators (${\rm OPO}_{i},{\rm OPO}_{ii}$) pumped below 
% oscillation threshold create two individual squeezed vacuum modes 
% $(\hat{x}_{i},\ \hat{p}_{i})$ and $(\hat{x}_{ii},\ \hat{p}_{ii})$. These 
% beams are first combined with a 50/50 beam splitter with a $\pi/2$ phase 
% shift and then one of the output beams is divided into two beams (B,C) 
% with another 50/50 beam splitter. The three output modes 
% $(\hat{x}_j,\ \hat{p}_j)\ (j={\rm A,B,C})$ (abbreviated as 
% $(\hat{x}_{\rm A,B,C},\ \hat{p}_{\rm A,B,C})$ hereafter) are entangled 
% with arbitrary levels of squeezing. 

% NEW VERSION:

Quantum telecloning relies on tripartite entanglement --- the minimum unit 
of multipartite entanglement. Tripartite entanglement for continuous variables 
can be generated by using squeezed vacuum states and two beam splitters 
\cite{vanLoock00}.  Even infinitesimal squeezing can yield fully 
inseparable tripartite state \cite{vanLoock03}.  The states so generated can 
be classified by the separability of the reduced bipartite state after 
tracing out one of the three subsystems. In the qubit regime, this 
classification is well established. For example, the 
Greenberger-Horne-Zeilinger (GHZ) state \cite{GHZ90} does not have any 
bipartite entanglement after the trace-out, while the $W$ state \cite{Duer00}
does. In the continuous-variable regime, various types of tripartite 
entanglement can be generated by choosing proper transmittances/reflectivities
of beam splitters and the levels of squeezing. For example, the 
continuous-variable analogue of the GHZ state \cite{vanLoock00,Yonezawa04}
was used in the quantum teleportation network. This state can be created 
by combining three squeezed vacuum states with a high level of squeezing 
on two beam splitters, and is a tripartite maximally entangled 
state in the limit of infinite squeezing. In the absence of any bipartite
entanglement between any pair of the three parties, quantum teleportation 
from a sender to a receiver cannot be achieved without the help of the 
third member. In contrast, the entanglement required for quantum telecloning 
comprises both a bipartite and a tripartite structure much like the $W$ 
state \cite{Duer00}, and it is not maximally entangled. We create this
new type of tripartite entanglement and use it to demonstrate $1 \to 2$ 
% new type of tripartite entanglement and use it in $1 \to 2$ 
quantum telecloning of coherent states. 
% The entanglement is generated using 
% two squeezed-vacuum states with a modest amount of squeezing and a pair of 
% 50/50 beam splitters.  

% We generate this 
% type of entanglement using two squeezed-vacuum states with a modest amount of 

% squeezing and a pair of 50/50 beam splitters. 

\begin{figure}[thb]
\begin{center}
\includegraphics[width=\linewidth]{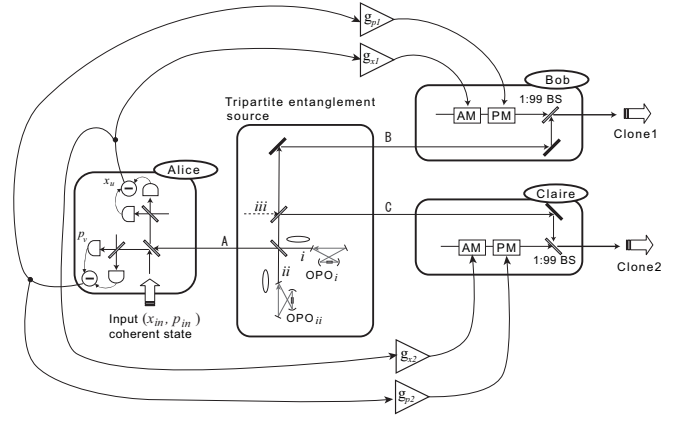}
\end{center}
\caption{The experimental set-up for quantum telecloning from Alice to 
Bob and Claire to produce Clone 1 and 2. }
\label{fig:setup}
\end{figure}

The scheme for creating the tripartite entanglement for quantum telecloning 
\cite{vanLoock01} is shown in the center of Fig.~\ref{fig:setup}. Two optical 
parametric oscillators (${\rm OPO}_{i},{\rm OPO}_{ii}$) pumped below 
oscillation threshold create two individual squeezed vacuum modes 
$(\hat{x}_{i},\ \hat{p}_{i})$ and $(\hat{x}_{ii},\ \hat{p}_{ii})$. These 
beams are first combined with a 50/50 beam splitter with a $\pi/2$ phase 
shift and then one of the output beams is divided into two beams (B,C) 
with another 50/50 beam splitter. The three output modes 
$(\hat{x}_j,\ \hat{p}_j)\ (j={\rm A,B,C})$ (abbreviated as 
$(\hat{x}_{\rm A,B,C},\ \hat{p}_{\rm A,B,C})$ hereafter) are entangled 
with arbitrary levels of squeezing.

% ----

Here, modes A and B and
modes A and C are bipartitely entangled and modes A, B, and C are
tripartitely entangled. On its own each mode is in a thermal state and
shows excess noises. This can be verified by applying the sufficient
inseparability criteria for a bipartite case \cite{Duan00,Simon00} and a
tripartite case \cite{vanLoock03}. In the present situation, the criteria
are
\begin{align}
\lefteqn{\langle [ \Delta(\hat{x}_{\rm A} - \hat{x}_{\rm B,C})]^2 \rangle
+
\langle [ \Delta(\hat{p}_{\rm A} + \hat{p}_{\rm B,C})]^2 \rangle} \nonumber \\
&=
\left ( \frac{1-\sqrt{2}}{2} \right )^2 
\left [ \langle ( \Delta \hat{x}_{i} )^2 \rangle 
+ \langle ( \Delta \hat{p}_{ii} )^2 \rangle \right ] \nonumber \\
+ &\left ( \frac{1+\sqrt{2}}{2} \right )^2
\left [ \langle ( \Delta \hat{x}_{ii} )^2 \rangle
+ \langle ( \Delta \hat{p}_{i} )^2 \rangle \right ] + \frac{1}{4}
%\nonumber \\
< 1\;,
\label{ineq:entangle}
\end{align}
where $\langle ( \Delta \hat{x}^{(0)} )^2 \rangle = \langle ( \Delta
\hat{p}^{(0)} )^2 \rangle =\frac{1}{4}$
and superscript ${(0)}$ denotes a vacuum.
The left-hand-side of the inequality can be minimized when $(\hat{x}_i,
\hat{p}_i)=(e^r \hat{x}_i^{(0)}, e^{-r} \hat{p}_i^{(0)})$, $(\hat{x}_{ii},
\hat{p}_{ii})=(e^{-r} \hat{x}_{ii}^{(0)}, e^{r} \hat{p}_{ii}^{(0)})$, and
$e^{-2r} = (\sqrt{2}-1)/(\sqrt{2}+1)$\ (7.7 dB squeezing).  By using these
tripartitely entangled modes, sender Alice can perform quantum telecloning
of a coherent state input to two receivers Bob and Claire to produce Clone
1 and 2 at their sites. In other words, success of quantum telecloning is
a sufficient condition for the existence of this type of entanglement.

For quantum telecloning, Alice first performs a joint measurement or
so-called ``Bell measurement'' on her entangled mode ($\hat{x}_{\rm A},
\hat{p}_{\rm A}$) and an unknown input mode ($\hat{x}_{\rm
in},\hat{p}_{\rm in}$). In our experiment, the input state is a coherent
state and a sideband of continuous wave 860 nm carrier light. The Bell
measurement instrument consists of a 50/50 beam splitter and two homodyne
detectors as shown in Fig.~\ref{fig:setup}. Two outputs of the input 50/50
beam splitter are labeled as $\hat{x}_u=(\hat{x}_{\rm in}-\hat{x}_{\rm
A})/\sqrt{2}$ and $\hat{p}_v=(\hat{p}_{\rm in} + \hat{p}_{\rm
A})/\sqrt{2}$ for the relevant quadratures. Before Alice's measurement,
the initial modes of Bob and Claire are, respectively,
\begin{eqnarray}
\hat{x}_{\rm B,C} &=& \hat{x}_{\rm in} - ( \hat{x}_{\rm A} - 
\hat{x}_{\rm B,C}) - \sqrt{2} \hat{x}_u
\\
\hat{p}_{\rm B,C} &=& \hat{p}_{\rm in} + ( \hat{p}_{\rm A} + 
\hat{p}_{\rm B,C}) - \sqrt{2} \hat{p}_v \;.
\end{eqnarray}
Note that in this step Bob's and Claire's modes remain unchanged. After
Alice's measurement on $\hat{x}_u$ and $\hat{p}_v$, these operators
collapse and reduce to certain values. Receiving these measurement results
from Alice, Bob and Claire displace their modes as $\hat{x}_{\rm B,C} \to
\hat{x}_{1,2}=\hat{x}_{\rm B,C}+\sqrt{2} x_u,\ \hat{p}_{\rm B,C} \to
\hat{p}_{1,2}=\hat{p}_{\rm B,C}+\sqrt{2} p_v$ and accomplish the
telecloning. Note that the values of $x_u$ and $p_v$ are classical
information and can be duplicated. In our experiment, displacement is
performed by applying electro-optical modulations. Bob and Claire modulate
beams by using amplitude and phase modulators (AM and PM in
Fig.~\ref{fig:setup}). The amplitude and phase modulations correspond to 
the displacement of $p$ and $x$ quadratures, respectively. The modulated 
beams are combined with Bob's and Claire's initial modes 
($\hat{x}_{\rm B,C}$, $\hat{p}_{\rm B,C}$) at $1/99$ beam splitters.

The output modes produced by the telecloning process are represented 
as \cite{vanLoock01},
\begin{eqnarray}
\hat{x}_{1,2} &=& \hat{x}_{\rm in} - ( \hat{x}_{\rm A} - \hat{x}_{\rm B,C})
\nonumber \\
&=& \hat{x}_{\rm in} + \frac{1-\sqrt{2}}{2} \hat{x}_i 
- \frac{1+\sqrt{2}}{2} \hat{x}_{ii} \pm \frac{1}{\sqrt{2}} \hat{x}_{iii}^{(0)}
\label{eq:x_12}
\\
\hat{p}_{1,2} &=& \hat{p}_{\rm in} + ( \hat{p}_{\rm A} + \hat{p}_{\rm B,C})
\nonumber \\
&=& \hat{p}_{\rm in} + \frac{1+\sqrt{2}}{2} \hat{p}_i 
- \frac{1-\sqrt{2}}{2} \hat{p}_{ii} \pm \frac{1}{\sqrt{2}} \hat{p}_{iii}^{(0)}
\label{eq:p_12} \;,
\end{eqnarray}
where subscript $iii$ denotes a vacuum input to the second beam splitter
in the tripartite entanglement source, and $+$ of $\pm$ for Clone 1 and
$-$ for Clone 2. From these equations, we can see that the telecloned
states have additional noise terms to the input mode ($\hat{x}_{\rm
in},\hat{p}_{\rm in}$). The additional noise can be minimized by tuning
the squeezing levels of the two output modes of the OPOs. This corresponds to
the minimization of the left-hand-side of Ineq. (\ref{ineq:entangle}).  In the
ideal case with 7.7 dB squeezing, the additional noise is minimized and we
obtain $\hat{x}_{1,2} = \hat{x}_{\rm in} -\frac{1}{2}(\hat{x}_i^{(0)} +
\hat{x}_{ii}^{(0)}) \pm \frac{1}{\sqrt{2}} \hat{x}_{iii}^{(0)}$ and
$\hat{p}_{1,2} = \hat{p}_{\rm in} +\frac{1}{2}(\hat{p}_i^{(0)} +
\hat{p}_{ii}^{(0)}) \pm \frac{1}{\sqrt{2}} \hat{p}_{iii}^{(0)}$. These are
the optimal clones of coherent state inputs \cite{vanLoock01}. In contrast
to quantum teleportation, these optimal clones are degraded from the
original input by one unit of vacuum noise. In the classical case, where
no quantum entanglement is used, two units of vacuum noise would be added.
This is dubbed the quduty which has to be paid for crossing the border
between quantum and classical domains \cite{Braunstein98}.

\begin{figure}[thb]
\begin{center}
\includegraphics[width=0.5\linewidth]{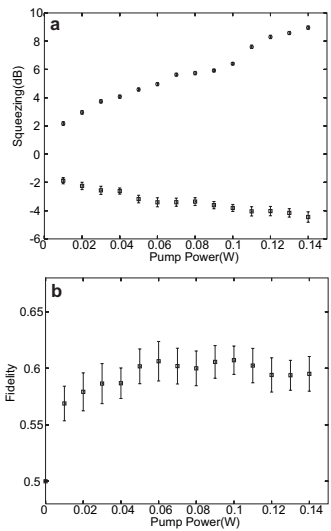}
\end{center}
\caption{(a) Pump power dependence of squeezing and antisqueezing of the
output of ${\rm OPO}_i$. The squeezing and antisqueezing are measured 
at 1 MHz. Visibility at a 50/50 beam splitter for homodyne measurement is 
about 0.95 and quantum efficiency of the detector is more than 99\%. (b) 
Calculated fidelities from the squeezing and antisqueezing.}
\label{fig:squeeze_fidelity}
\end{figure}

% In the real experiments, however, the output modes of OPOs suffer from
% inevitable losses and deviate from the ideal case. Particularly, the
% mixedness of the states makes the experimental procedure more complicated
% than the above discussion based on pure states. Now the states are not
% minimum uncertainty states, but show $\langle ( \Delta \hat{x}_{i,ii} )^2
% \rangle \langle ( \Delta \hat{p}_{i,ii} )^2 \rangle > 1/16$. The added
% noise gives rise to the asymmetry between squeezing and antisqueezing. 
% An appropriate choice of two parameters for the squeezing and antisqueezing
% allows us to minimize this additional noise.
% This is again in contrast to quantum teleportation: in principle, the 
% antisqueezing terms are canceled out and the asymmetry never affects the 
% performance of teleportation. 

To evaluate the performance of telecloning, we use the fidelity 
$F = \langle \psi_{\rm in} | \hat{\rho}_{\rm out} | \psi_{\rm
in} \rangle$ \cite{Braunstein00,Braunstein01}. The classical limit for 
coherent state cloning is derived by averaging the fidelity for a
randomly chosen coherent input 
$F_{\rm av}=\frac{1}{2}$ \cite{Braunstein00,Hammerer05}. Experimentally,
it is impossible to average over the entire phase space. However, if the 
gains of the classical channels 
$g_{x1,x2}= \langle \hat{x}_{1,2} \rangle / \langle \hat{x}_{\rm in}
\rangle$ and $g_{p1,p2}= \langle \hat{p}_{1,2} \rangle / \langle
\hat{p}_{\rm in} \rangle$ are unity $g_{x1,x2}=g_{p1,p2 }=1$, the averaged
fidelity is identical to the fidelity for a particular coherent state
input ($F_{\rm av} = F$). This is because the fidelity with unity gains
is fully determined by the variances of the telecloned states, 
independent of the amplitude of the coherent state input. Experimental
adjustment of $g_x=g_p=1$ is performed in the manner of
Ref.~\onlinecite{Zhang03}. The fidelity for a coherent state input with 
$g_x=g_p=1$ can be written as \cite{Furusawa98},
\begin{equation}
F = 2/ \sqrt{[1+4 \langle ( \Delta \hat{x}_{1,2} )^2 \rangle] [1+4 
\langle ( \Delta \hat{p}_{1,2} )^2 \rangle]}\;.
\label{eq:F_with_var}
\end{equation}
From the above discussion, if we measure $\langle ( \Delta \hat{x}_{1,2}
)^2 \rangle$ and $\langle ( \Delta \hat{p}_{1,2} )^2 \rangle$ of the
outputs for a coherent state input and get $F > \frac{1}{2}$, then 
the quantum telecloning of coherent states is deemed successful. 
Note that the optimal fidelity of Gaussian coherent-state telecloning 
\cite{vanLoock01} is $\frac{2}{3}$, which is consistent with the 
parameters of the ideal case mentioned above (see Ref.~\onlinecite{cerf05}
for a non-Gaussian result).

% By using Eq.~(\ref{eq:F_with_var}) and experimental results of
% squeezing/antisqueezing, we calculate the expected fidelities of the
% telecloning experiments. 

Fig.~\ref{fig:squeeze_fidelity}a shows the
typical pump power dependence of squeezing and antisqueezing of the
output of the OPOs. Here the OPO cavities contain Potassium Niobate
crystals inside as nonlinear mediums and are pumped with the frequency 
doubled outputs of a continuous wave Ti:sapphire laser at 860 nm. In 
order to minimize the asymmetry of squeezing without sacrificing the 
level of squeezing, we select mirrors with reflectivity of 12\% for the output
couplers of the OPOs. With Eqs.~(\ref{eq:x_12}), (\ref{eq:p_12}) and
(\ref{eq:F_with_var}) and these experimental results, we calculate the 
expected fidelities of the telecloning experiments, which are plotted in 
Fig.~\ref{fig:squeeze_fidelity}b. Accordingly, we set the pump power to 60 mW 
% for which we expect the best fidelity to be $\simeq 0.6$.
for which we expect the fidelity to be $\simeq 0.6$.

\begin{figure}[thb]
\begin{center}
\includegraphics[width=3.0truein]{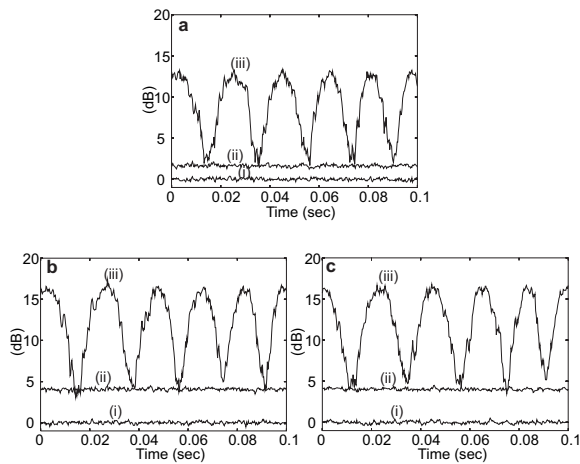}
\end{center}
\caption{Quantum telecloning from Alice to Bob and Claire. All traces are
normalized to the corresponding vacuum noise levels. (a) Alice's
measurement results for $p$ quadrature.
% ($x$ quadrature is not shown).
Trace i, the corresponding vacuum noise level $\langle ( \Delta
\hat{p}^{(0)} )^2 \rangle =\frac{1}{4}$. Trace ii, the measurement result
of a vacuum input $\langle ( \hat{p}_v^{\prime} )^2 \rangle$ where
$\hat{p}_v^{\prime}=(\hat{p}_{\rm in}^{(0)}+\hat{p}_{\rm A})/\sqrt{2}$.
Trace iii, the measurement result of a coherent state input $\langle (
\hat{p}_v )^2 \rangle$ with the phase scanned.  (b),(c) The measurement
results of the telecloned states at Bob (b) and Claire (c) for $p$
quadratures ($x$ quadratures are not shown). Trace i, the corresponding
vacuum noise levels.  Trace ii, the telecloned states for a vacuum input
$\langle ( \Delta \hat{p}_{1,2} )^2 \rangle$.  Trace iii, the telecloned
states for a coherent state input.  The measurement frequency is centered
at 1 MHz, and the resolution and video bandwidths are 30 kHz and 300 Hz, 
respectively. All traces except for trace iii are averaged 20 times.}
\label{fig:result_coherent}
\end{figure}

Quantum telecloning was performed for two types of input states: a vacuum 
state and a coherent state that is created by applying electro-optic 
modulation to a very weak carrier beam.  Fig.~\ref{fig:result_coherent}
summarizes the results from both experiments, with Alice's states 
in Fig.~\ref{fig:result_coherent}a, and Bob and Claire's output states
in Fig.~\ref{fig:result_coherent}b and c. Trace ii of 
Fig.~\ref{fig:result_coherent}a shows Alice's $p$-quadrature measurement 
for a vacuum input, $\langle ( \hat{p}_v^{\prime} )^2 \rangle$, where
$\hat{p}_v^{\prime}=(\hat{p} _{\rm in}^{(0)}+\hat{p}_{\rm A})/\sqrt{2}$.
Note that $\langle \hat{p}_{\rm in}^{(0)} \rangle = \langle \hat{p}_{\rm
A} \rangle=0$; thus $\langle ( \hat{p}_v^{\prime} )^2 \rangle= \langle (
\Delta \hat{p}_v^{\prime} )^2 \rangle = 
\langle ( \Delta \hat{p}_v )^2 \rangle$, because the vacuum is a 
zero-amplitude coherent state.  The noise level is $2.1$ dB higher
compared to the vacuum noise level
$\langle ( \Delta \hat{p}^{(0)} )^2 \rangle =\frac{1}{4}$, 
due to the ``entangled noise'' $\hat{p}_{\rm A}$. This noise is canceled 
to some extent by the tripartite entanglement. Trace iii in 
Fig.~\ref{fig:result_coherent}a shows Alice's coherent-state input with 
the phase scanned. Consistent with the above discussion on the variance 
of a coherent state input, the troughs of trace iii are level with 
trace ii within experimental accuracy. Note that Alice's 50/50 beam 
splitter reduces the amplitude of the measured state (i.e., the peaks of 
trace iii) by 3 dB relative to the input state.

Figs.~\ref{fig:result_coherent}b and c show the measurement results of the 
telecloned states. Traces ii show the results for a vacuum input, 
$\langle ( \Delta \hat{p}_{1,2} )^2 \rangle$.  The noise level for Clone 1 
is $4. 06 \pm 0.17$ dB and that for Clone 2 is $4.03 \pm 0.15$ dB. We also 
measured the $x$-quadratures
$\langle ( \Delta \hat{x}_{1,2} )^2 \rangle$ and obtained $3.74 \pm 0.15$ dB 
for Clone 1 and $3.79 \pm 0.15$ dB for Clone 2 (not shown). 
Note that the telecloned states have the same mean amplitude as that of 
the input inferred from Alice's measurement, which is consistent with the 
unit gains of the classical channels. Finally, we calculated the fidelity
from Eq.~(\ref{eq:F_with_var}), and found $F = 0.58 \pm 0.01$ for both
teleclones.  Since this fidelity exceeds the classical cloning limit of 
$\frac{1}{2}$, we have successfully demonstrated $1 \to 2$ quantum 
telecloning of coherent states. These results are operational evidence 
for the existence of the tripartite entanglement. The slight 
discrepancies from the expected fidelities are attributed to fluctuations
of phase locking of the system.

We note that, in quantum cryptographic scenarios, quantum telecloning, 
complemented by quantum storage, may provide a means for an eavesdropper 
to remotely monitor a quantum cryptographic channel more securely. In
particular, the remote nature of the operation offers the advantage that 
even if eavesdropping is discovered, her identity and location 
are guaranteed uncompromised. As distinct from the symmetric telecloning
reported here, asymmetric telecloning might be the method of choice by a 
technologically advanced eavesdropper. This can easily be achieved in our
scheme by modifying the shared state and feedforward gains. In addition, 
to make the scheme practical, high 
capacity quantum memory would be essential for the multiply entangled 
states used to operate the protocol autonomously. This would guarantee 
the eavesdropper's anonymity and the flexibility to perform individual,
collective or coherent attacks without the need for backward 
communication. Promising strides in the storage of continuous-variables 
states \cite{Lukin,Polzik} would also give suitable storage for entangled 
states. Thus, the 
successful demonstration of coherent-state telecloning offers a step 
forward in the technological toolkit of eavesdroppers of quantum 
cryptographic channels.

In conclusion, we have demonstrated $1 \to 2$ quantum telecloning of
coherent states for continuous variables.  Manipulations of multipartite 
entanglement are essential for realization of quantum computation and 
quantum communication among many parties.  In particular, this paper
reported a demonstration of manipulations of a new type of multipartite 
entanglement and an example of the reduction of the number of steps in 
quantum information processing with entanglement. The techniques used 
in this experiment are easily extendable to $ 1 \to n$ quantum telecloning 
and related operations.

%\begin{acknowledgments}
This work was partly supported by the MEXT and the MPHPT of Japan. SLB 
currently holds a Wolfson - Royal Society Research Merit award. The 
authors appreciated discussions with Netta Cohen and Peter van Loock.
%\end{acknowledgments}

\end{document}